\documentclass[aps,pre,twocolumn,floats,showpacs]{revtex4}

\usepackage{epsfig}
\usepackage{color}
\usepackage{bm}
\usepackage{latexsym}

\begin{document}
\newcommand{\hide}[1]{}
\newcommand{\tbox}[1]{\mbox{\tiny #1}}
\newcommand{\half}{\mbox{\small $\frac{1}{2}$}}
\newcommand{\sinc}{\mbox{sinc}}
\newcommand{\const}{\mbox{const}}
\newcommand{\trc}{\mbox{trace}}
\newcommand{\intt}{\int\!\!\!\!\int }
\newcommand{\ointt}{\int\!\!\!\!\int\!\!\!\!\!\circ\ }
\newcommand{\eexp}{\mbox{e}^}
\newcommand{\bra}{\left\langle}
\newcommand{\ket}{\right\rangle}
\newcommand{\EPS} {\mbox{\LARGE $\epsilon$}}
\newcommand{\ar}{\mathsf r}
\newcommand{\im}{\mbox{Im}}
\newcommand{\re}{\mbox{Re}}
\newcommand{\bmsf}[1]{\bm{\mathsf{#1}}}
\newcommand{\mpg}[2][1.0\hsize]{\begin{minipage}[b]{#1}{#2}\end{minipage}}

\title{Universality in the spectral and eigenfunction properties of random networks}

\author{J. A. M\'endez-Berm\'udez}
\email{jmendezb@ifuap.buap.mx}
\affiliation{Instituto de F\'{\i}sica, Benem\'erita Universidad Aut\'onoma de Puebla,
Apartado Postal J-48, Puebla 72570, Mexico}

\author{A. Alcazar-L\'opez}
\affiliation{Instituto de F\'{\i}sica, Benem\'erita Universidad Aut\'onoma de Puebla,
Apartado Postal J-48, Puebla 72570, Mexico}

\author{A. J. Mart\'inez-Mendoza}
\affiliation{Instituto de F\'{\i}sica, Benem\'erita Universidad Aut\'onoma de Puebla,
Apartado Postal J-48, Puebla 72570, Mexico}
\affiliation{Elm\'eleti Fizika Tansz\'ek, Fizikai Int\'ezet, Budapesti M\H uszaki \'es Gazdas\'agtudom\'anyi
Egyetem, H-1521 Budapest, Hungary}

\author{Francisco A. Rodrigues}
\affiliation{Departamento de Matem\'{a}tica Aplicada e Estat\'{i}stica, Instituto de Ci\^{e}ncias
Matem\'{a}ticas e de Computa\c{c}\~{a}o, Universidade de S\~{a}o Paulo, Caixa Postal 668,13560-970
S\~{a}o Carlos,  S\~ao Paulo, Brazil}

\author{Thomas K. DM. Peron}
\affiliation{Instituto de F\'{\i}sica de S\~{a}o Carlos, Universidade de S\~{a}o Paulo, S\~{a}o Carlos,
S\~ao Paulo, Brazil}

\date{\today}

\begin{abstract}
By the use of extensive numerical simulations we show that the nearest-neighbor energy
level spacing distribution $P(s)$ and the entropic eigenfunction localization
length of the adjacency matrices of Erd\H{o}s-R\'enyi (ER) {\it fully} random networks
are universal for fixed average degree $\xi\equiv \alpha N$ ($\alpha$ and $N$ being
the average network connectivity and the network size, respectively).
We also demonstrate that Brody distribution characterizes well $P(s)$ in the transition from
$\alpha=0$, when the vertices in the network are isolated, to $\alpha=1$, when the network
is fully connected. Moreover, we explore the validity of our findings when relaxing
the randomness of our network model and show that, in contrast to standard ER networks,
ER networks with {\it diagonal disorder} also show universality. Finally, we also discuss
the spectral and eigenfunction properties of small-world networks.
\end{abstract}


\maketitle


\section{Introduction}

Networks have been used to represent the organization of complex systems, such as social networks,
Internet, and ecosystems \cite{Costa011:AP,B13}. Depending on the application, vertices and edges
have different meanings \cite{Costa011:AP,N10}. For example, in condensed matter physics, vertices
and edges of the ordered network known as Anderson's tight-binding model are the sites and hopping
integrals, respectively \cite{3DAM}. Networks can be deterministic, fractal, or random \cite{Mulken2011}.
Deterministic and fractal networks are constructed following specific rules, whereas for random
networks a set of parameters take fixed values but the network itself has a random organization.
In this later case it is meaningless to study a single random network, instead a statistical analysis
of an ensemble of networks with the same average properties should be performed. Several models
of random networks have been introduced \cite{N10,Boccaletti06}, including Erd\H{o}s-R\'{e}nyi (ER)
random graphs, the scale-free network model of Barab\'{a}si and Albert, and the small-world
networks of Watts and Strogatz. These models are considered to reproduce the organization of
real-world networks, such as the Internet, power-grids, and social and biological networks
\cite{N10, Boccaletti06, Costa011:AP}. Although random graphs fail in predicting most of the
properties observed in real-world networks, such as power-law degree distributions and
nonvanishing clustering coefficient \cite{Boccaletti06}, such graphs have been deeply studied
theoretically (e.g.~\cite{Bollobas98}). Indeed, many results, such as the emergence of percolation,
can be obtained analytically in ER networks \cite{N10,Bollobas98}.

Here we consider the ER random graph model, which was introduced by Solomonoff and Rapoport \cite{SR51}
and deeply studied later by Erd\H{o}s and R\'enyi \cite{ER59,ER60}.
This model is also known as uncorrelated random graph model. ER networks are constructed by
starting with $N$ isolated vertices and afterwards, each pair of vertices is
connected according to a probability $\alpha$. This process is a type of $N^2$-realizations Bernoulli
process with probability of success $\alpha$. Therefore, the number of connections follows a binomial
distribution. Nevertheless, most realizations of this model take into account large values of $N$
and small values of $\alpha$. In this way, the degree distribution tends to a Poisson distribution due
to the law of rare events.

Independently of the field, classification, or application, a commonly accepted mathematical representation
of a network is the adjacency matrix. The adjacency matrix $\bf A$ of a simple network, i.e. a
network having no multiple edges or self-edges, is the matrix with elements $A_{ij}$ defined as
\cite{N10}
\begin{equation}
A_{ij}=\left\{
\begin{array}{ll}
1 \ &\text{if there is an edge between vertices $i$ and $j$},\\
0 \ &\text{otherwise}.
\end{array}
\right.
\label{Aij}
\end{equation}
This prescription produces $N\times N$ symmetric sparse matrices with zero diagonal elements, where
$N$ is the number of vertices of the corresponding network. The sparsity of $\bf A$ is quantified by the
parameter $\alpha$ which is the fraction of non-vanishing off-diagonal adjacency matrix elements.
Vertices are isolated when $\alpha=0$, whereas the network is fully connected for $\alpha=1$. Once the
adjacency matrix of a network is constructed, it is quite natural to ask about its spectral and eigenfunction
properties, which is the main subject of this paper. As commonly used, we refer to the spectral and
eigenfunction properties of the adjacency matrix as the spectral and eigenfunction properties of the
respective network.

Moreover, there is a one-to-one correspondence between the adjacency matrix $\bf A$ and the Hamiltonian matrix
$\bf H$ of a $\xi$-dimensional solid, described by Anderson's tight-binding model \cite{3DAM} with zero
on-site potentials ($H_{ii}=0$) and constant hopping integrals ($H_{ij}=1$). Here $\xi$ is proportional
to the average non-zero off-diagonal adjacency matrix elements per matrix row and, therefore, may be
regarded as the effective dimension of the network represented by $\bf A$; as discussed in \cite{JMR01}
from a random matrix theory (RMT) point of view. This correspondence enables the direct application of
studies originally motivated on physical systems, represented by Hamiltonian sparse random matrices, to
complex networks. In the network literature, $\xi$ is known as the average degree of a network.
Moreover, notice that
\begin{equation}
\label{xi}
\xi = \alpha \times N \ ,
\end{equation}
where $\xi$ was defined as the mean number of nonzero elements per matrix row. From a mathematical-physicist
point of view, in the frame of RMT, in 1988 Rodgers and Bray \cite{RB88} proposed an ensemble of sparse random
matrices characterized by the connectivity $\xi$. Since then, several papers have been devoted to
analytical and numerical studies of sparse symmetric random matrices (see for example
\cite{JMR01,RB88,RD90,MF91,EE92,E92,SC02,KR97,FM91b,K08,RPK08,S11,KTW10,RP09,F14}).

Among the most relevant results of these studies we can mention that:
(i) in the very sparse limit, $\xi\to 1$, the density of states was found to deviate from the Wigner semicircle
law with the appearance of singularities, around and at  the band center, and tails beyond the semicircle
\cite{RB88,RD90,MF91,EE92,E92,SC02,KR97,K08,RPK08,S11};
(ii) a delocalization transition was found at $\xi\approx 1.4$ \cite{MF91,FM91b,EE92,E92};
(iii) the nearest-neighbor energy level spacing distribution $P(s)$ was found to evolve from the Poisson to the
Gaussian Orthogonal Ensemble (GOE) predictions for increasing $\xi$ \cite{EE92,E92,JMR01} (the same transition
was reported for the number variance in Ref.~\cite{JMR01}). More recently, the first eigenvalue/eigenfunction
problem was also addressed in Ref.~\cite{KTW10}. Also, non-Hermitian sparse matrices were approached in
Refs.~\cite{RP09}.

It is relevant to emphasize that the RMT model of sparse matrices introduced by Rodgers and Bray \cite{RB88}
is equivalent to adjacency matrices of ER--type networks. In fact, motivated by this equivalency and based on
the ER model, here we study spectral and eigenfunction properties of the following random network model:
Starting with the {\it standard} ER network, we add to it self-edges and further consider all edges to have
random strengths \cite{noteA}. We call this model as the ER {\it fully} random network model. The sparsity
$\alpha$ is defined as the fraction of the $N(N-1)/2$ independent non-vanishing off-diagonal adjacency matrix
elements. Then, as in the {\it standard} ER model, the ER {\it fully} random network model is characterized
only by the parameters $N$ and $\alpha$. However, the corresponding adjacency matrices come from the ensemble
of $N\times N$ sparse real symmetric matrices whose non-vanishing elements are statistically independent random
variables drawn from a normal distribution with zero mean $\bra A_{ij} \ket=0$ and variance
$\bra |A_{ij}|^2 \ket=(1+\delta_{ij})/2$. According to this definition a diagonal random matrix is obtained for
$\alpha=0$ (Poisson case), whereas the GOE is recovered when $\alpha=1$.

Our motivation to study spectral and eigenfunction properties of the ER {\it fully} random network model is
twofold. On the one hand, Jackson et al.~\cite{JMR01} showed that the disorder parameter $\xi$ fixes some
spectral features of sparse random matrices equivalent to the adjacency matrices we study here. Moreover,
it is also known that the average degree $\xi$ fixes some properties of random graphs \cite{AB02}. On the
other hand, there is a large number of papers studying spectral and eigenfunction properties
of complex networks
\cite{PV06,DR93,ZX00,GGS05,SKHB05,JKBH08,BJ07,JB08,ZYYL08,J09,JB07,GT06,ZX01,GKK01,F01,F02,DGMS03,AM05,KC05,RAKK05,MPV07,NR08,JB09,EK09,CJH09,CAS08,GGS09,JSVL10,JZL11,S12}.
However, most of those studies focus on networks with specific combinations of $N$ and $\alpha$.
For instance Refs.~\cite{JKBH08,BJ07,JB08,ZYYL08,J09,JB07,JB09} investigated random networks using concepts of
RMT centering the attention on networks with fixed $N$ and $\alpha$, even when more sophisticated topological
properties are included in the network models, such as non-vanishing clustering coefficient~\cite{JKBH08} and
modular structure~\cite{J09}.
Moreover, Palla and Vallay~\cite{PV06} investigated spectral properties of {\it standard} ER networks close to
the critical point of percolation restricting the analysis to values of $1<\xi<2.4$.

Then, we realized that the universality (in the sense of identifying relevant network parameters) of the
spectral and eigenfunction properties of ER--type random networks has not been completely explored yet.
Thus, in this paper we undertake this task having as a reference RMT models and predictions.
Furthermore, our investigations generalize previous results in the literature by not restricting our analysis
to a specific combination of network parameters or even to a specific network model.

In the next section we analyze $P(s)$, i.e. the nearest-neighbor energy level spacing distribution, the average
Shannon entropy $\bra S \ket$, and the entropic eigenfunction localization length $\ell_N$ for the ER {\it
fully} random network  model as a function of the connectivity $\alpha$. It is well known that $P(s)$, the
probability distribution function of the spacings between adjacent eigenvalues, plays a prominent role in the
description of disordered and quantized classically-chaotic systems \cite{metha,Haake10}. We
show that $P(s)$, $\bra S \ket$, and $\ell_N$ are all invariant for fixed average degree $\xi$. In addition, we
show that Brody distribution fits well $P(s)$ in the transition from isolated to fully connected networks. By
noticing that the ER {\it fully} random network model displays maximal disorder, in Sect.~III  we explore
the validity of our findings when relaxing the randomness of  this network model. Therefore we show that the ER
network model with {\it diagonal disorder} (i.e. ER model including random strength self-edges) also exhibits
universality for fixed $\xi$. We also apply our approach to a different random network model, so in
Sect.~IV we comment on the spectral and eigenfunction properties of small-world random networks.
Our conclusions are developed in Sect.~V.

\section{Erd\H{o}s-R\'enyi fully random networks}

In the following we use exact numerical diagonalization to obtain the eigenvalues $E^m$ and eigenfunctions
$\Psi^m$ ($m=1\ldots N$) of the adjacency matrices of large ensembles of random networks characterized by $N$
and $\alpha$.

\subsection{Nearest-neighbor energy level spacing distribution}

\begin{figure}[t]
    \centering
    \includegraphics[width=0.9\columnwidth]{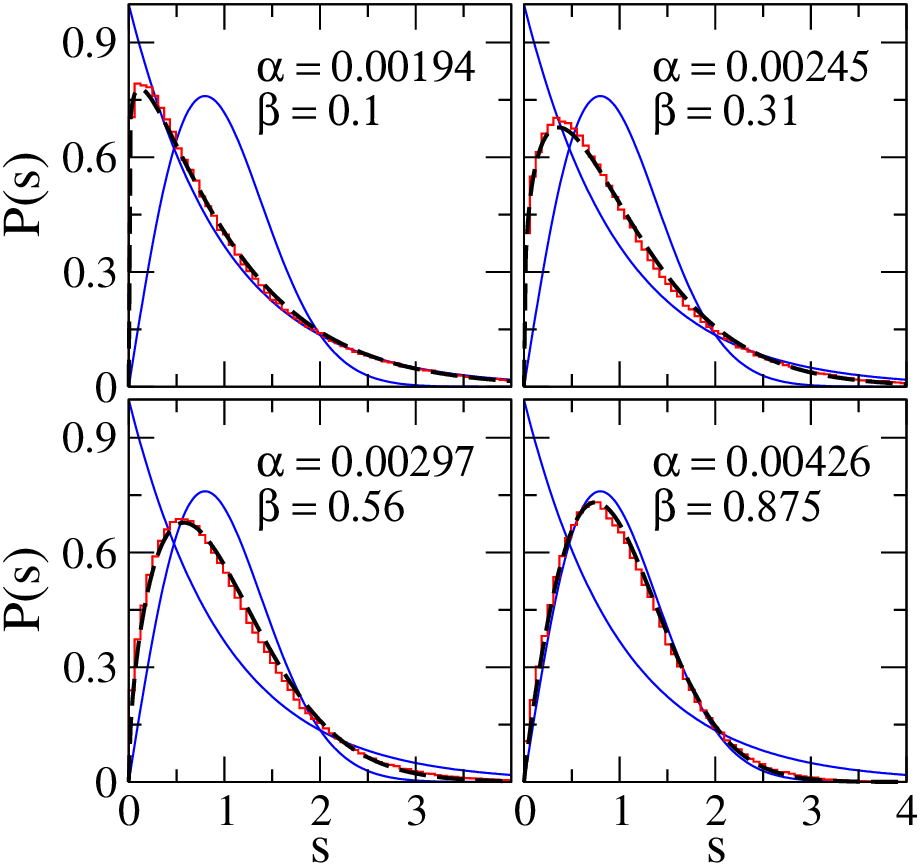}
\caption{(Color online) Nearest-neighbor energy level spacing distribution $P(s)$ for ER {\it fully} random
networks of size $N=1000$ and different connectivity values $\alpha$ (red histograms). Blue lines correspond
to Poisson and Wigner-Dyson distribution functions given by Eqs.~(\ref{P}) and (\ref{WD}), respectively.
Black dashed lines are fittings of the histograms with the Brody distribution of Eq.~(\ref{B}), where the
fitted values of $\beta$ are given in the corresponding panels. The histograms were computed from $5\times10^5$
unfolded spacings.}
\label{Fig1}
\end{figure}

Figure~\ref{Fig1} presents the nearest-neighbor energy level spacing distribution $P(s)$ for the adjacency
matrices of ER {\it fully} random networks of size $N=1000$ and different connectivity values $\alpha$. The
histograms were constructed by the use of the 500 unfolded spacings \cite{metha},
$s_m=(E^{m+1}-E^m)/\Delta$, around the band center of $10^3$ random matrices. Here, $\Delta$ is the mean level
spacing computed for each adjacency matrix as the slope of the curve $E^m$ vs.~$m$ around the band center.
When other network sizes are considered, we always construct $P(s)$ from half of the total eigenvalues around
the band center, where the density of states is approximately constant.

For $\alpha=0$, i.e. when the vertices in the network are isolated, the corresponding adjacency matrices are
diagonal and $P(s)$ follows the exponential distribution,
\begin{equation}
\label{P}
P(s) = \exp(-s) \ ,
\end{equation}
better known in RMT as Poisson distribution or the spacing rule for random levels \cite{metha}.
In the opposite limit, $\alpha=1$, when the network is fully connected, the adjacency matrices become members
of the GOE (full real symmetric random matrices) and $P(s)$ closely follows the Wigner-Dyson
distribution~\cite{metha,note0},
\begin{equation}
\label{WD}
P(s) = \frac{\pi}{2} s \exp \left(- \frac{\pi}{4} s^2 \right) \ .
\end{equation}
Then, by increasing $\alpha$ from zero to one, the shape of $P(s)$ should evolve from the Poisson to the
Wigner-Dyson distributions. This transition, partially observed for our random network model in
Refs.~\cite{EE92,JMR01} from a pure RMT point of view, is well depicted in Fig.~\ref{Fig1}, where we also plot
Eqs.~(\ref{P}) and (\ref{WD}) as reference. In fact, it is interesting to mention that for relatively small
values of $\alpha$ the limiting GOE statistics is recovered. The transition from Poisson to Wigner-Dyson in the
spectral statistics has also been reported for adjacency matrices corresponding to other complex network
models in Refs.~\cite{ZX00,GGS05,SKHB05,JKBH08,BJ07,JB08,ZYYL08,J09}.

Here, in order to characterize the shape of $P(s)$ for our random networks we use the Brody distribution
\cite{B73,B81} which was originally derived to provide an interpolation expression for $P(s)$ in the transition
from Poisson to Wigner-Dyson distributions by making the ansatz
$P(s) = c_1s^\beta\exp\left(-c_2s^{\beta+1}\right)$ (with
$c_{1,2}$ depending on $\beta$). Then, after proper normalization Brody distribution reads \cite{B73,B81}
\begin{equation}
\label{B}
P(s) = (\beta +1) a_{\beta} s^{\beta}  \exp\left(- a_{\beta} s^{\beta+1}\right) \ ,
\end{equation}
where $a_{\beta} = [\Gamma(\beta+2/\beta+1)]^{\beta+1}$, $\Gamma(\cdot)$ is the gamma function, and $\beta$,
known as Brody parameter, takes values in the range $[0,1]$. $\beta=0$ and $\beta=1$ reproduce the Poisson and
Wigner-Dyson distributions, respectively. We want to remark that, even though Brody distribution has been
extensively used to characterize $P(s)$ having fractional power-law level repulsion \cite{examples}, it has
been obtained through a purely phenomenological
approach and the Brody parameter has no decisive physical meaning \cite{physical} but serves as a measure for the
degree of mixing between the Poisson and GOE statistics.
In particular, as we show below, the Brody parameter will allows us to identify the onset of the delocalization
transition and the onset of the GOE limit in our random network models.

Black dashed lines in Fig.~\ref{Fig1} represent the fittings of the
numerically obtained $P(s)$ with Eq.~(\ref{B}). The fitted values of $\beta$ are given in the corresponding
panels. This figure shows that Brody distribution provides very good fittings for the $P(s)$ of the adjacency
matrices of ER {\it fully} random networks. In fact, Brody distribution also works well for other complex
networks models \cite{BJ07,JB08,ZYYL08,J09,JB07}.

\begin{figure}[t]
    \centering
    \includegraphics[width=0.9\columnwidth]{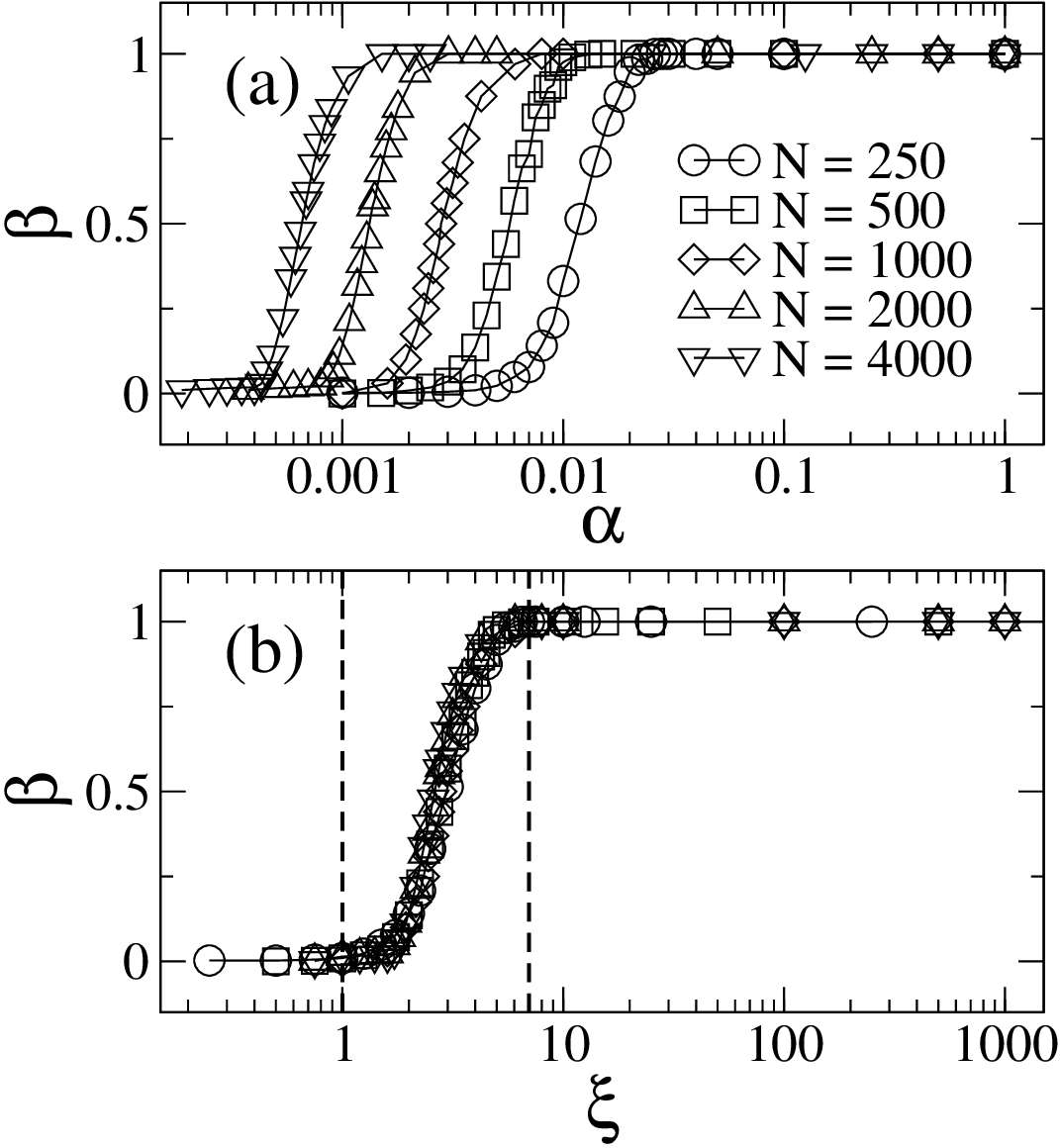}
\caption{Brody parameter $\beta$ as a function of (a) the connectivity $\alpha$ and (b) the average degree
$\xi=\alpha N$ for ER {\it fully} random networks of sizes ranging from $N=250$ to 4000. Dashed vertical lines
at $\xi=1$ and $\xi=7$ mark the onset of the delocalization transition and the onset of the GOE limit,
respectively. Error bars are not shown since they are much smaller than symbol size.}
\label{Fig2}
\end{figure}

Now we also construct histograms of $P(s)$ for a large number of values of $\alpha$ to extract systematically
the corresponding values of $\beta$. Figure~\ref{Fig2}(a) reports $\beta$ versus $\alpha$ for five different
network sizes. Notice that in all five cases the behavior of $\beta$ is similar: $\beta$ shows a smooth
transition from zero (Poisson regime) to one (Wigner-Dyson or GOE regime) when $\alpha$ increases from
$\alpha\ll 1$ (mostly isolated vertices) to one (fully connected networks). Notice also that the larger the
network size $N$, the smaller the value of $\alpha$ needed to approach the GOE limit. This is, in fact, the
reason we observed in Fig.~\ref{Fig1} that $P(s)$ is very close to the Wigner-Dyson distribution already for
$\alpha=0.00426\ll 1$.

\begin{figure}[t]
    \centering
    \includegraphics[width=0.9\columnwidth]{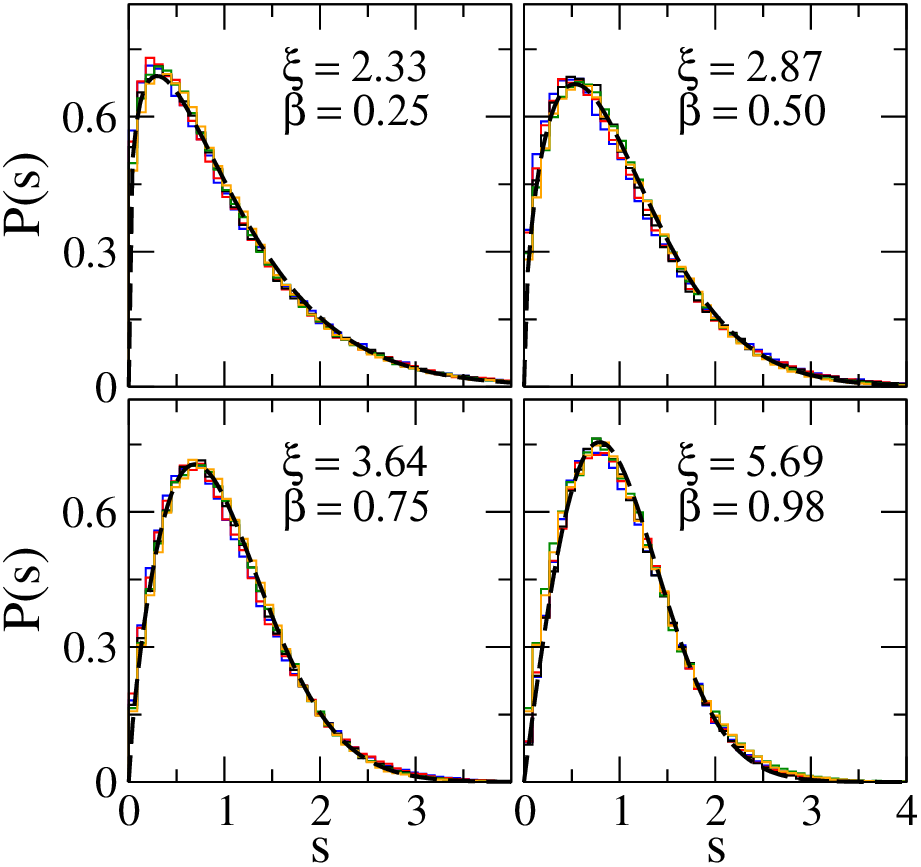}
\caption{(Color online) Nearest-neighbor energy level spacing distribution $P(s)$ for ER {\it fully} random
networks of sizes ranging from $N=250$ to 4000 and different values of the average degree $\xi$ (color
histograms). Black dashed lines are the Brody distribution of Eq.~(\ref{B}) with $\beta=0.25$, 0.5, 0.75, and
0.98. Each histogram was computed from $5\times10^5$ unfolded spacings.}
\label{Fig3}
\end{figure}
\begin{figure}[t]
    \centering
    \includegraphics[width=0.9\columnwidth]{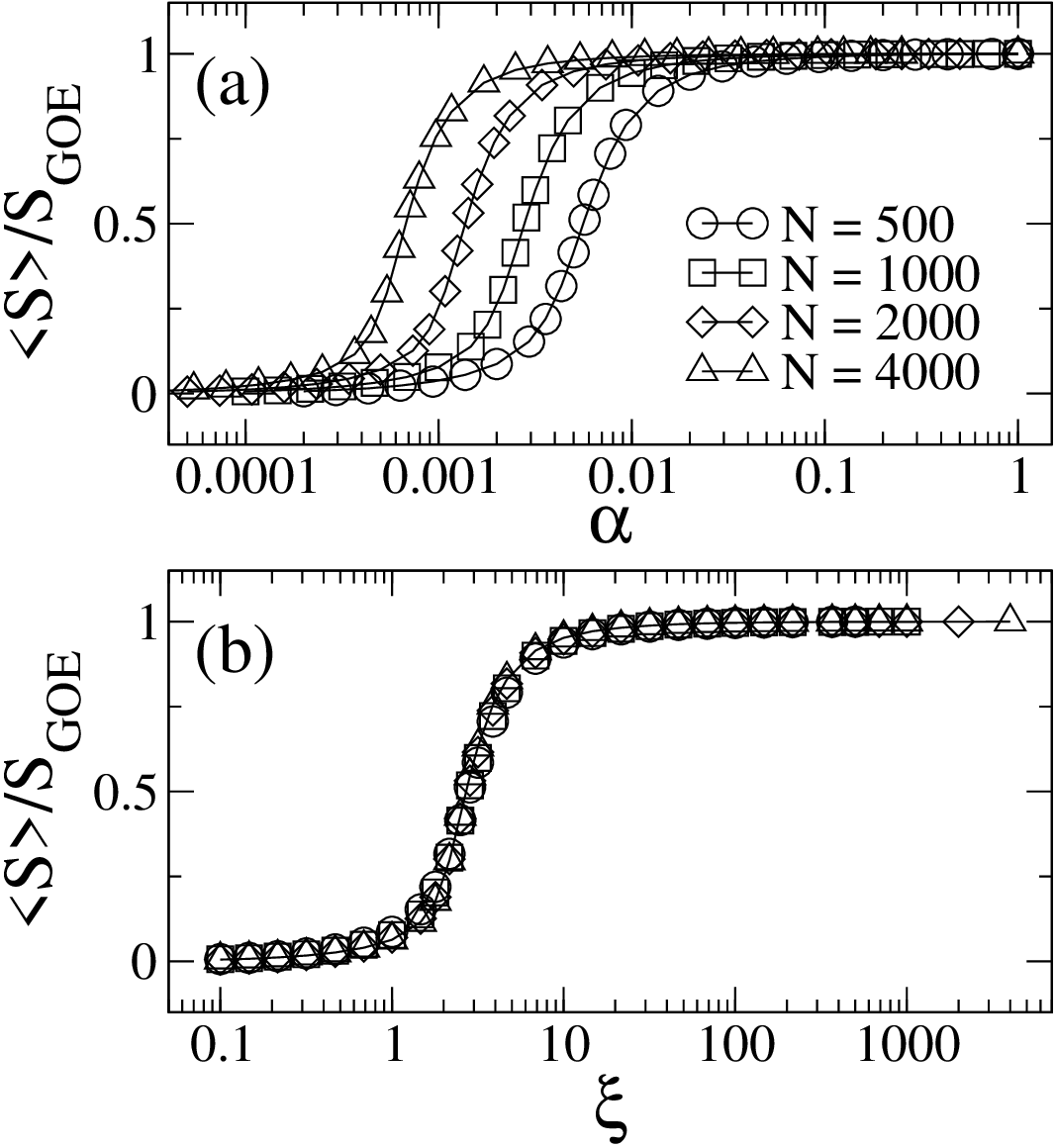}
\caption{Average Shannon entropy $\bra S \ket$ normalized to $S_{\tbox{GOE}}$ as a function of (a) the
connectivity $\alpha$ and (b) the average degree $\xi$ for ER {\it fully} random networks of sizes ranging from
$N=500$ to 4000. Each point was computed by averaging over $10^6$ eigenfunctions.}
\label{Fig4}
\end{figure}

We recall that the parameter $\xi$ (see Eq.~(\ref{xi})) fixes the density of states of sparse random
matrices~\cite{JMR01} equivalent to the density of connections in the adjacency matrices we study here. Moreover,
$\xi$ was defined as the effective dimension of the sparse random matrix model~\cite{JMR01}. Also,
Mart\'inez-Mendoza et al.~\cite{MAM13} showed that scattering and transport properties of tight-binding ER {\it
fully} random  networks are universal for fixed disorder parameter $\xi$. Thus, it make sense to
explore the dependence of $\beta$ on the average degree $\xi$; in fact, in Fig.~\ref{Fig2}(b) we show this 
dependency. We observe that curves for different network sizes $N$ fall on top of a
universal curve. This means that once $\xi$ is fixed, no matter the network size, the shape of $P(s)$ is also
fixed. We also note that the transition in the form of $P(s)$ takes place in the interval $1<\xi<7$; i.e.
when $\xi\le 1$ ($\xi\ge 7$), $P(s)$ has the Poisson (Wigner-Dyson) shape.

Therefore, we verify the invariance of the form of $P(s)$ for fixed average degree $\xi$ by (i) choosing four
representative values of $\beta$ (0.25, 0.5, 0.75, and $0.98\sim 1$);  (ii) extracting the corresponding values
of $\xi$ from the universal curve of Fig.~\ref{Fig2}(b); and (iii) constructing histograms of $P(s)$ for
several network sizes for each of the chosen values of $\xi$. Figure~\ref{Fig3} shows that once $\xi$ is fixed,
the form of $P(s)$ is invariant. In addition, we include in each panel the corresponding Brody distributions.

Notice that Fig.~\ref{Fig2}(b) also provides a way to predict the shape of $P(s)$ of ER {\it fully} random
networks once the average degree $\xi$ is known:  When $\xi<1$, $P(s)$ has the Poisson shape. For $\xi>7$,
$P(s)$ is practically given by the Wigner-Dyson distribution. While in the regime $1\le\xi\le 7$, $P(s)$ is
well described by Brody distributions characterized by a value of $0<\beta<1$. Thus, $\xi=1$ and $\xi=7$ mark
the onset of the delocalization transition and the onset of the GOE limit, respectively.

\subsection{Entropic eigenfunction localization length}

In order to characterize quantitatively the complexity of the eigenfunctions of random matrices (and of
Hamiltonians corresponding to disordered and quantized chaotic systems) two quantities are mostly used: (i) the
information or Shannon entropy and (ii) the eigenfunction participation number. These measures provide the
number of principal components of an eigenfunction in a given basis. In fact, both quantities have been already
used to characterize the eigenfunctions of the adjacency matrices of random network models (see some examples in
Refs.~\cite{F01,MPV07,GT06,ZYYL08,GGS05,CAS08,GGS09,JSVL10,JZL11,S12,PS12,PMCM13,MRP13,MZN14}).

Here, we use the Shannon entropy, which for the eigenfunction $\Psi^m$ is given as
\begin{equation}
\label{S}
S = -\sum_{n=1}^N (\Psi^m_n)^2 \ln (\Psi^m_n)^2 \ .
\end{equation}
$S$ allows us to compute the so called entropic eigenfunction localization length \cite{I90}, i.e.
\begin{equation}
\label{lH}
\ell_N = N \exp\left[ -\left( S_{\tbox{GOE}} - \bra S \ket \right)\right] \ ,
\end{equation}
where $S_{\tbox{GOE}}\approx\ln(N/2.07)$ is the entropy of a random eigenfunction with Gaussian distributed
amplitudes. We average over all eigenfunctions of an ensemble of adjacency matrices of size $N$ to compute
$\bra S \ket$~\cite{note1}. With this definition, when $\alpha=0$, since the eigenfunctions of the adjacency
matrices of our random network model have only one non-vanishing component with magnitude equal to one,
$\bra S \ket=0$ and $\ell_N\approx 2.07$. On the other hand, for $\alpha=1$, $\bra S \ket=S_{\tbox{GOE}}$ and
the fully chaotic eigenfunctions extend over the $N$ available vertices in the network, i.e.
$\ell_N\approx N$.

\begin{figure}[t]
    \centering
    \includegraphics[width=0.9\columnwidth]{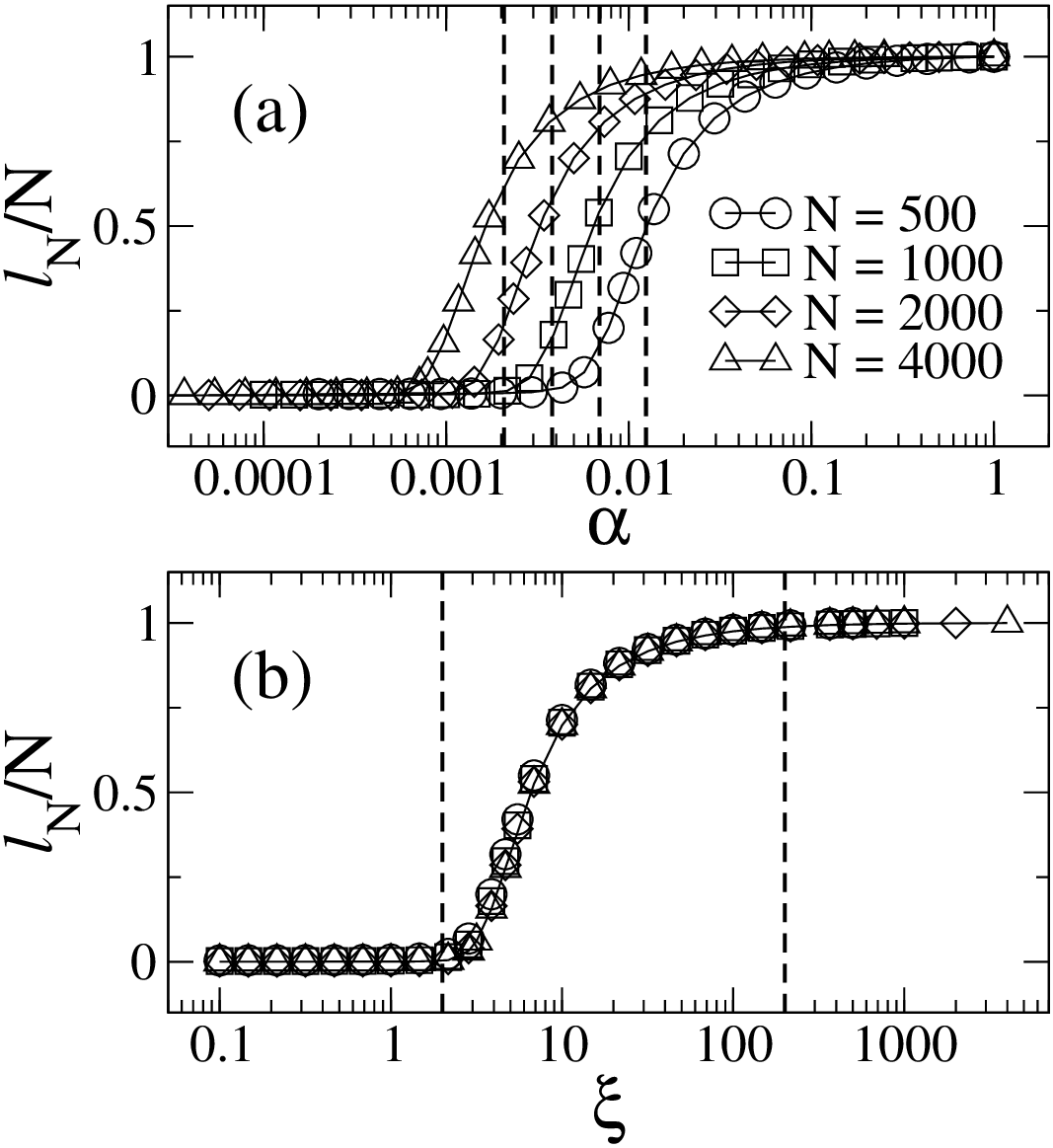}
\caption{Entropic eigenfunction localization length $\ell_N$ normalized to $N$ as a function of (a) the
connectivity $\alpha$ and (b) the average degree $\xi$ for ER {\it fully} random networks of sizes ranging
from $N=500$ to 4000. Dashed vertical lines in (a) mark the values of $\alpha=(\ln N)/N$ at which every ER
{\it fully} random network becomes globally connected \cite{AB02}. Dashed vertical lines in (b) at $\xi=2$
and $\xi=200$ mark the onset of the delocalization transition and the onset of the GOE limit, respectively.
Each point was computed by averaging over $10^6$ eigenfunctions.}
\label{Fig5}
\end{figure}

Figures~\ref{Fig4}(a) and \ref{Fig5}(a) show $\bra S \ket/S_{\tbox{GOE}}$ and $\ell_N/N$, respectively, as a
function of the connectivity $\alpha$ for the adjacency matrices of ER {\it fully} random networks of sizes
$N=500$, 1000, 2000, and 4000. We observe that the curves $\bra S \ket/S_{\tbox{GOE}}$ and $\ell_N/N$ have the
same functional form as a function of $\alpha$. Notice that this behavior is also observed for $\beta$ in
Fig.~\ref{Fig2}(a). Also these curves are displaced to the left for increasing $N$. On the other hand, when
we plot $\bra S \ket/S_{\tbox{GOE}}$ and $\ell_N/N$ as a function of the average degree $\xi$, we observe the
coalescence of all curves into universal ones (see Figs.~\ref{Fig4}(b) and \ref{Fig5}(b)). In addition, note
that the point at which every ER {\it fully} random network becomes globally connected (in the sense that it
does not contain isolated sub-networks),  $\alpha=(\ln N)/N$ \cite{AB02}, occurs when $\ell_N/N\approx 1/2$
(see the dashed vertical lines in Fig.~\ref{Fig5}(a)).

From Fig.~\ref{Fig5}(b) it is clear that the universal behavior of the curve $\ell_N/N$ as a function of the
average degree $\xi$ can be easily described:
(i) $\ell_N/N$ transits from $\approx 2.07/N\sim 0$ to one by moving $\xi$ from zero to $N$;
(ii) for $\xi\stackrel{<}{\sim}2$ the eigenfunctions are practically localized since $\ell_N\sim 1$; hence the
delocalization transition takes place around $\xi\approx 2$, which is close to previous theoretical and
numerical estimations \cite{MF91,FM91b,EE92,E92}; and
(iii) for $\xi>200$ the eigenfunctions are practically chaotic and fully extended since
$\ell_N \approx N$. Despite the fact that the transition region for $\ell_N/N$, which takes place in the range
$2<\xi<200$, is very large as compared to that for $\beta$, both quantities are highly correlated and
characterize well the Poisson to Wigner-Dyson transition of spectral and eigenfunction properties of the
adjacency matrices of ER {\it fully} random networks as a function of $\xi$.

\section{Other Erd\H{o}s-R\'enyi random networks}

Notice that with the prescription given above our ER {\it fully} random network model displays maximal
disorder, because averaging over the network ensemble implies average over connectivity and over connection
strengths. With this averaging procedure we get rid off any individual network characteristic (such as scars
\cite{SK03}, which in turn produce topological resonances \cite{GSS13}) that may lead to deviations from
RMT predictions used here as a reference (see also \cite{PJRK13}). More specifically, we
choose this network model to retrieve well known random matrices in the appropriate limits --- remember that a
diagonal random matrix is obtained for $\alpha=0$, when the vertices are isolated, whereas a member of the
GOE is recovered for $\alpha=1$, when the network is fully connected.

However, it is important to add that the maximal disorder we consider above is not necessary for a
graph/network to exhibit universal RMT behavior. In fact, it is well known that tight-binding cubic lattices
with on-site disorder (known as the three-dimensional Anderson model \cite{3DAM}), forming networks with fixed
regular connectivity having very dilute adjacency matrices, show RMT behavior in the metallic phase (see
for example Refs.~\cite{metallic1,metallic2}). Moreover, it has been demonstrated numerically and theoretically
that graphs with fixed connectivity show spectral~\cite{spectral,TS01} and scattering \cite{PW13,scattering}
universal properties corresponding to RMT predictions. In this case the disorder is introduced either by
choosing random bond lengths \cite{spectral,PW13,scattering} (which is a parameter not present in our network
model) or by randomizing the vertex-scattering matrices \cite{TS01} (somehow equivalent to consider random
connection strengths). Some of the RMT properties of quantum graphs have already been tested experimentally by
the use of small ensembles of small microwave networks with fixed connectivity \cite{Sirko}. Furthermore, complex
networks having specific topological properties (such as small-world and scale-free networks, where randomness
is applied only to the connectivity) show signatures of RMT behavior in their spectral and eigenfunction
properties \cite{BJ07,JB08,ZYYL08,JB07,MPV07}.

Therefore, in the following we search for the scaling properties, if any, of ER--type random networks when the
condition of maximal disorder considered above is relaxed.

\subsection{Standard Erd\H{o}s-R\'enyi random networks}

In the {\it standard} ER random network model \cite{SR51,ER59,ER60}, the corresponding adjacency matrices are
random matrices with zeros in the main diagonal and ones as non-vanishing off-diagonal elements; i.e. in the
adjacency matrix vertices and edges are represented with zeros and ones, respectively (see Eq.~(\ref{Aij})).
Even though it has been shown that the $P(s)$ of {\it standard} ER random networks is close to the Wigner-Dyson
shape for large connectivity ($\alpha\to 1$) \cite{BJ07,JB08,JB07}, notice that $P(s)$ can not show the Poisson
to Wigner-Dyson transition since in the limit of vanishing connectivity ($\alpha=0$) the corresponding adjacency
matrices are the null matrix.

\begin{figure}[t]
    \centering
    \includegraphics[width=0.9\columnwidth]{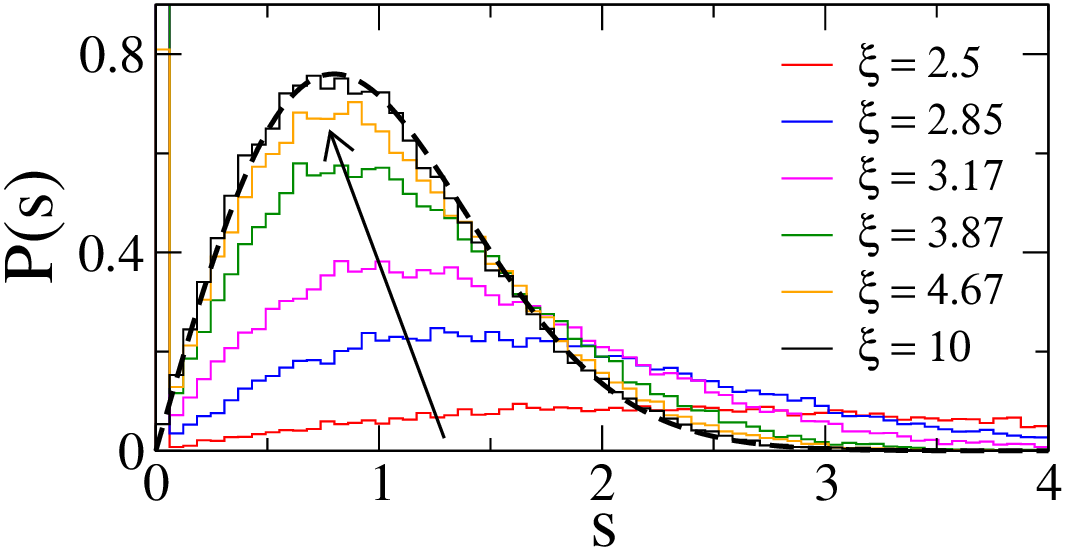}
\caption{(Color online) Nearest-neighbor energy level spacing distribution $P(s)$ for {\it standard} ER
random networks of size $N=1000$ and different values of the average degree $\xi$ (color histograms). Black
dashed line is the Wigner-Dyson distribution of Eq.~(\ref{WD}). The arrow indicates the direction of increasing
$\xi$. Each histogram was computed from $5\times10^5$ unfolded spacings.}
\label{Fig6}
\end{figure}

Figure~\ref{Fig6} shows the distribution $P(s)$ for {\it standard} ER random networks of size $N=1000$. We
observe that once $\xi$ is large enough $P(s)$ acquires the expected Wigner-Dyson shape. However, for smaller
values of $\xi$, $P(s)$ develops two components: (i) a prominent peak at $s=0$ and (ii) a broad part having a
local maximum at $s>0$. The presence of these two components avoids the use of the Brody distribution to
describe the shape of $P(s)$ in the transition from isolated vertices to fully connected networks. The same
behavior is observed for any other value of $N$.

It is fair to say that the Brody parameter has already been reported as a function of $\xi$ for {\it standard} 
ER random networks in Ref.~\cite{PV06}; however, for $1 \le \xi \le 2.4$ only. The reason to report
such a narrow range of $\xi$ is because for vanishing connectivity this model does not approach the Poisson
limit and $P(s)$ can not be fitted by Brody distribution.

Concerning the Shannon entropy and the entropic eigenfunction localization length for this random network model,
we observe, as expected, that $\bra S \ket/S_{\tbox{GOE}}$ and $\ell_N/N$ approach zero when $\alpha\to 0$,
whereas they approach one when $\alpha\to 1$. Thus, the delocalization transition of this network model can be
well characterized by both $\bra S \ket$ and $\ell_N$. However, there is no universal scaling of
$\bra S \ket/S_{\tbox{GOE}}$ and  $\ell_N/N$ when plotted as a function of $\xi$ (not shown here).

\subsection{Erd\H{o}s-R\'enyi random networks with diagonal disorder}

\begin{figure}[t]
    \centering
    \includegraphics[width=0.9\columnwidth]{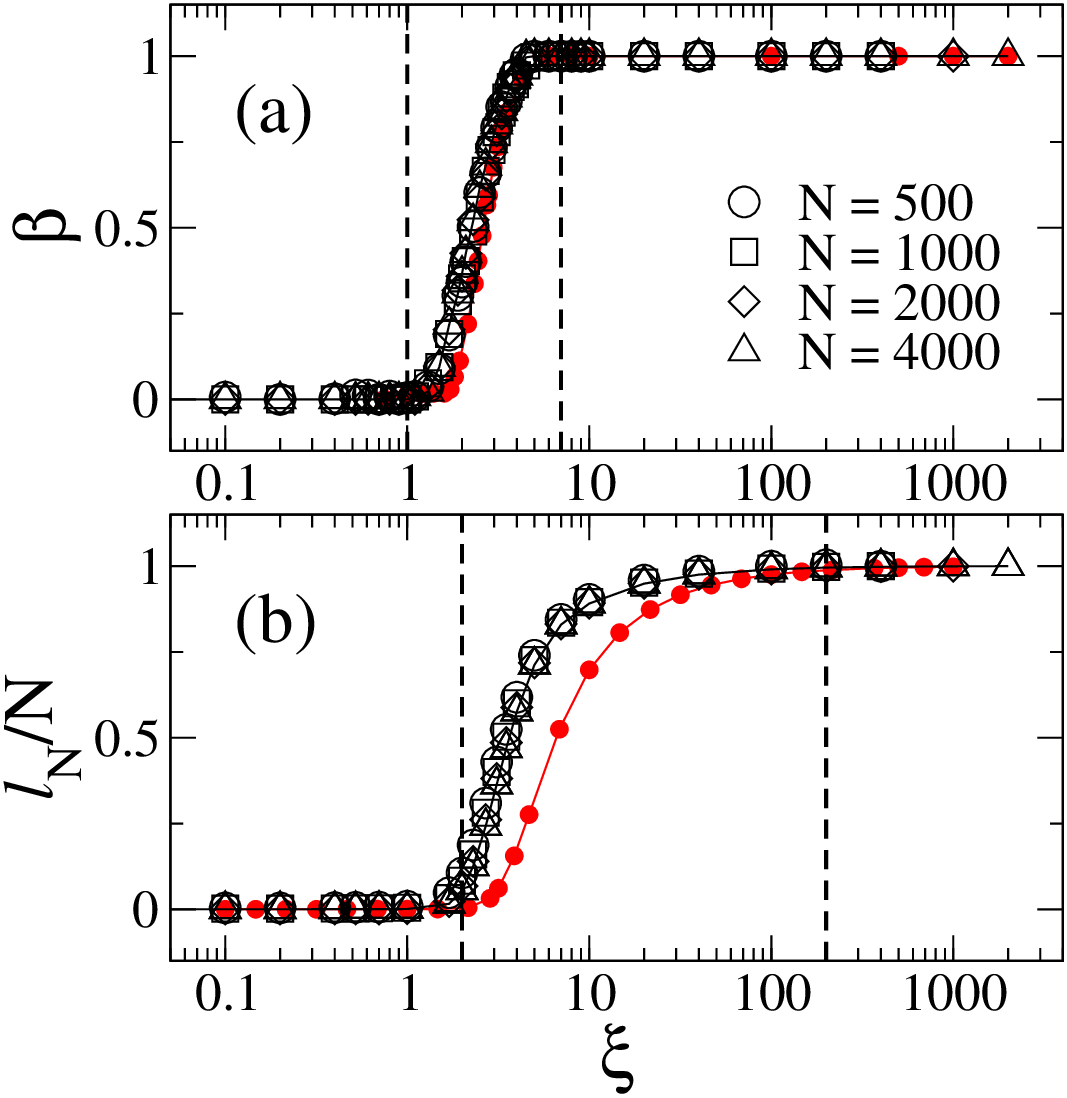}
\caption{(Color online) (a) Brody parameter $\beta$ and (b) entropic eigenfunction localization length $\ell_N$
(normalized to $N$) as a function of the average degree $\xi$ for ER random networks with {\it diagonal
disorder} of sizes ranging from $N=500$ to 4000. Dashed red vertical lines, displayed for comparison purposes,
at $\xi=1$ and $\xi=7$ in (a) [at $\xi=2$ and $\xi=200$ in (b)] mark the onset of the delocalization transition
and the onset of the GOE limit, respectively, of ER {\it fully} random networks; see Fig.~\ref{Fig2}(b) [see
Fig.~\ref{Fig5}(b)]. Red data sets corresponding to ER {\it fully} random networks of size $N=4000$ are also
included as a reference.}
\label{Fig7}
\end{figure}

We construct the ER random network model with {\it diagonal disorder} by adding self-edges with random strengths such
that the corresponding adjacency matrices acquire statistically independent random variables (drawn from a
normal distribution with zero mean and variance one) in their main diagonal \cite{note3}. With this
construction, we do observe a clear Poisson to Wigner-Dyson transition in the form of $P(s)$ when $\alpha$ moves
from zero to one. Moreover, $P(s)$ can be effectively fitted by the Brody distribution
(not shown here). Thus, Fig.~\ref{Fig7}(a) presents the Brody parameter $\beta$ as a function of the average
degree $\xi$ for ER random networks with {\it diagonal disorder}. As well as for ER {\it fully} random networks,
here we observe that the curves of $\beta$ vs.~$\xi$ for different network sizes $N$ fall on top of a
universal curve. In addition, we also observe universal behavior for the entropic eigenfunction localization
length $\ell_N$, normalized to $N$, as a function of $\xi$ (see Fig.~\ref{Fig7}(b)).
The delocalization transition in ER networks with {\it diagonal disorder} has
also been investigated in Ref.~\cite{SKHB05}, but as a function of the disorder strength, by the use
of $P(s)$ for networks having $3 \le \xi \le 10$.

It is relevant to add that even though we observe the collapse of the curves $\beta$ vs.~$\xi$ and $\ell_N/N$
vs.~$\xi$ for ER random networks with {\it diagonal disorder} of different sizes, these universal curves are
slightly displaced to the left as compared to the corresponding curves for ER {\it fully} random networks. More
specifically, the onset of the delocalization transition and the onset of the GOE limit occurs for smaller
values of $\xi$ in the case of ER random networks with {\it diagonal disorder}. See dashed lines and red curves
in Fig.~\ref{Fig7} (included for comparison purposes) which correspond to ER {\it fully} random networks.

\section{Small-world networks}

Once we analyzed the universal properties of ER--type random networks, it make sense to further explore other
random networks models to look for universal properties. For this task we consider small-world (SW)
networks~\cite{WS98}. A SW network, as defined in 1998 by Watts and Strogatz \cite{WS98}, is constructed by
randomly rewiring the edges of a regular ring network consisting of $N$ vertices connected to their $k/2$
nearest-neighbors ($k\ge 2$ must be an even number) \cite{note4}. Then, for every vertex, every right-handed
edge is reconnected with probability $p$ to a vertex chosen uniformly at random. In the {\it standard} SW model
multiply connected vertex pairs and self-connections  are not allowed. Notice that for $p=0$ the SW network
becomes the original regular ring; whereas for $p=1$, a random network is obtained where every vertex has a
minimum degree of $k/2$.

Notice that: (i) The parameter $k$ is in fact equivalent to the average degree $\xi$, given in (\ref{xi}), which
fixes the spectral and eigenfunction properties of ER {\it fully} random networks and ER random networks with
{\it diagonal disorder}. However, $k$ cannot take non-integer values nor values less than two as $\xi$ does.
Also, it is not an average quantity. (ii) The rewiring probability $p$ is a parameter, independent of $k$, which
drives the SW network model from regular to random. Also, we should stress that as well as for {\it standard} ER
random networks, the adjacency matrices of {\it standard} SW networks have zeros in their main diagonal. This
prohibits the use of the Brody distribution to fit the $P(s)$ of {\it standard} SW networks when $p$ and $k$ are
both small. However, for large $p$ and $k$, $P(s)$ becomes close to the Wigner-Dyson distribution, as
shown in Refs.~\cite{BJ07,ZYYL08,JB07,JB09} (for an analytical approach to the spectra of SW networks see
Ref.~\cite{timme}).

Then, as we did in the previous Section for ER random networks, here we also consider SW networks with {\it
diagonal disorder}. That is, we include self-edges having random strengths (drawn from a normal distribution
with zero mean and variance one) to each vertex in the network, such that the corresponding adjacency matrices
exhibit random variables in their main diagonal, whereas the edges joining vertex pairs are still represented by
ones in the adjacency matrices. In this way we guarantee that $P(s)$ will have the
Poisson shape when $p$ and $k$ are both small. Moreover, we observe (not shown here) that by including
{\it diagonal disorder}, the Brody distribution fits reasonably well the $P(s)$ of SW networks for any
combination of $p$ and $k$. The delocalization transition in SW networks with {\it diagonal disorder} has
also been investigated in Ref.~\cite{GGS05} as a function of $p$, $N$, and disorder strength by the use
of the inverse participation ratio of eigenfunctions (equivalent to the entropic eigenfunction
localization length we use here); however, the goal there was the implementation of a quantum algorithm
able to efficiently simulate the network.

\begin{figure}[t]
    \centering
    \includegraphics[width=0.9\columnwidth]{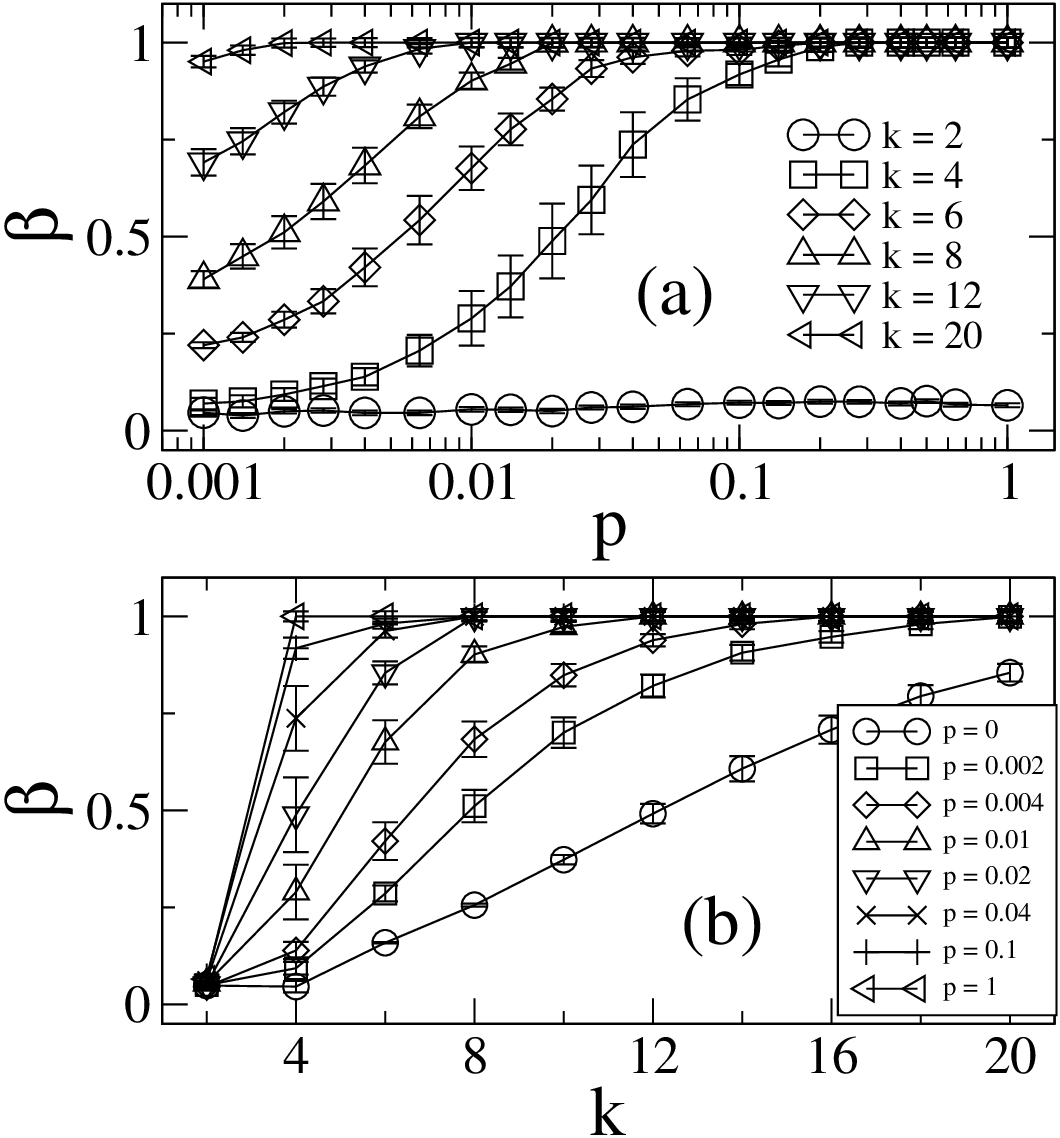}
\caption{Brody parameter $\beta$ as a function of (a) the rewiring probability $p$ and (b) the degree $k$ for
SW networks with {\it diagonal disorder} of size $N=1000$.}
\label{Fig8}
\end{figure}

Figure~\ref{Fig8} shows the Brody parameter $\beta$ as a function of $p$ (for several values of $k$) and $k$
(for several values of $p$) for SW networks with {\it diagonal disorder} of size $N=1000$. Except for $k=2$,
where $P(s)$ is very close to Poissonian for any $p$ (i.e. $\beta\approx 0$), since the corresponding adjacency
matrix is almost diagonal, $P(s)$ shows the transition from Poisson to Wigner-Dyson as a function of $p$ (see
Fig.~\ref{Fig8}(a)). From this figure we can see that the larger the value of $k$, the smaller the value of $p$
needed for $\beta$ to approach one. Notice, for example, that for $k=20$ the value of $p=0.001\ll 1$ makes
$\beta$ to be very close to one. Clearly, the transition from Poisson to Wigner-Dyson in the form of $P(s)$ is
also observed as a function of $k$ for fixed $p$ (see Fig.~\ref{Fig8}(b)).

\begin{figure}[t]
    \centering
    \includegraphics[width=0.9\columnwidth]{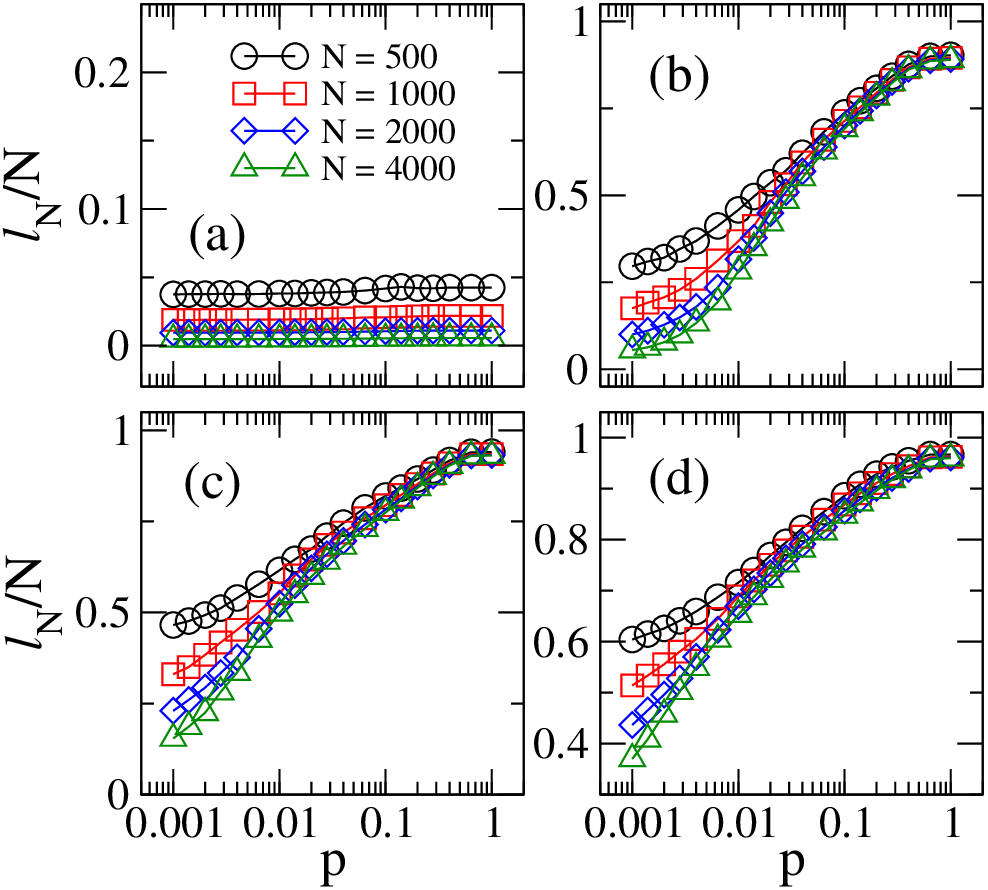}
\caption{(Color online) Entropic eigenfunction localization length $\ell_N$ (normalized to $N$) as a function of
the rewiring probability $p$ for SW networks with {\it diagonal disorder} of sizes ranging from $N=500$ to 4000.
(a) $k=2$, (b) $k=8$, (c) $k=12$, and (d) $k=20$.}
\label{Fig9}
\end{figure}

Also, in Ref.~\cite{BJ07} the Brody parameter has been reported as function of $p$ but for {\it standard} 
SW networks with fixed size $N = 2000$ and average degree $k=40$. The goal there was to show that $\beta$
is correlated with two important network parameters: the characteristic path length and the clustering
coefficient; quantities which are out of the scope of our paper. Anyhow, the delocalization transition
was clearly shown as a function of $p$ (analogous to our curves reported in Fig.~\ref{Fig8}(a) for SW 
networks with {\it diagonal disorder}).

We have to stress that, in contrast to ER random networks with {\it diagonal disorder}, where the average degree
$\xi$ fixes their spectral and eigenfunction properties, the degree $k$ does not fix the properties of SW
networks with {\it diagonal disorder}. This fact is clearly visible in Fig.~\ref{Fig9}, where we show $\ell_N/N$
as a function of $p$ for SW networks of four different sizes. Each panel corresponds to a fixed value of degree
$k$ where the curves approach each other only when $p>0.1$.

Finally, we have to mention that in analogy with the ER {\it fully} random networks we studied in Sect.~II, here
we also considered the case of SW {\it fully} random networks; i.e. all non-zero entries of the corresponding
adjacency matrices, including the main diagonal, were considered as random variables (in fact, this case has been
approached in Ref.~\cite{ZX00} where the transition from Poisson to Wigner-Dyson in the shape of $P(s)$ was 
reported as a function of $p$ for a network with the fixed parameters $N=1600$ and $k=8$).
However, we have not observed a substantial difference to the results reported in Figs.~\ref{Fig8} and
\ref{Fig9} (not shown here).

\section{Conclusions}

We have studied numerically some spectral and eigenfunction properties of Erd\H{o}s-R\'enyi--type random
networks focusing our attention on universality.

In particular, we have shown for Erd\H{o}s-R\'enyi (ER) {\it fully} random networks (where all non-vanishing
adjacency matrix elements are Gaussian random variables) and ER networks with {\it diagonal disorder} (where the
diagonal adjacency matrix elements are Gaussian random variables, whereas the rest of non-vanishing matrix
elements are ones) that: (i) The nearest-neighbor energy level spacing distribution $P(s)$, the average Shannon
entropy $\bra S \ket$, and the entropic eigenfunction localization length $\ell_N$ are universal for fixed
average degree $\xi=\alpha N$ (where $\alpha$ and $N$ are the network connectivity and the network size,
respectively); (ii) Brody distribution fits well $P(s)$ in the transition from Poisson ($\alpha=0$; isolated
vertices) to Wigner-Dyson ($\alpha=1$; fully connected network); and (iii) the Brody parameter $\beta$ as a
function of $\xi$ displays an invariant curve.

This analysis provides a way to predict the shape of $P(s)$ of ER--type random networks once the parameter
$\xi$ is known.

Specifically, for ER {\it fully} random networks and ER networks with {\it diagonal disorder} we have found that
when $0<\xi<1$, $P(s)$ is well described by the Poisson shape. This range of $\xi$ values coincides with the
regime where a typical ER random graph is composed of isolated trees \cite{AB02}. We have heuristically located
the delocalization transition point around $\xi\approx 1$, in close agreement with the transition value of
$\xi\approx 1.4$ reported in Refs.~\cite{MF91,FM91b,EE92,E92}. Also note that $\xi\approx 3.5$, known as the
average degree value at which the diameter of an ER random graph equals the diameter of the giant cluster
\cite{AB02}, is located about half the way in the Poisson to Wigner-Dyson transition. Also, we have observed
that for $P(s)$ to approach the Wigner-Dyson shape $\xi\ge 7$ is needed.

However, we have determined that there is no universal scaling of $\bra S \ket/S_{\tbox{GOE}}$ and $\ell_N/N$
when plotted as a function of $\xi$ for the {\it standard} ER model nor for SW networks; even though these two
quantities describe well the delocalization transition of both models. Moreover, we can affirm that {\it
diagonal disorder} is needed for a random network model to show the Poisson to Wigner-Dyson transition in the
form of $P(s)$.

\begin{acknowledgments}

This work was partially supported by VIEP-BUAP (Grant No.~MEBJ-EXC15-I), Fondo Institucional PIFCA 
(Grant No.~BUAP-CA-169), and CONACyT (Grant Nos.~I0010-2014-02/246246 and CB-2013-01/220624).
FAR acknowledges CNPq (Grant No.~305940/2010-4), FAPESP (Grant No.~2011/50761-2 and 2013/26416-9) 
and NAP eScience - PRP - USP for financial support. TKMP would like to acknowledge FAPESP (Grant 
No.~2012/22160-7) for the scholarship provided.

\end{acknowledgments}


\begin{thebibliography}{99}
\bibitem{noteA}
We justify the addition of random strengths to edges by recognizing
that in real-world networks connections and interactions are in general
non-constant.
\bibitem{note1}
We have verified that our observations are not modified when we
restrict the averages to a fraction of the eigenfunctions around
the band center.
\bibitem{note3}
This random network model also corresponds to tight-binding random
networks with random on-site potentials and constant (normalized to
one) coupling amplitudes between sites; which is in fact the standard
prescription for condensed matter models of the Anderson type.
\bibitem{note4}
Note that when $k=2$, the ring is a one-dimensional thigh-binding ring.
\end{thebibliography}

\bibliographystyle{plainnat}

\end{document}